# FedMinds: Privacy-Preserving Personalized Brain Visual Decoding

GUANGYIN BAO

Tongji University, College of Electronic and Information Engineering, baogy@tongji.edu.cn

DUOQIAN MIAO*

Tongji University, College of Electronic and Information Engineering, dqmiao@tongji.edu.cn

Exploring the mysteries of the human brain is a long-term research topic in neuroscience. With the help of deep learning, decoding visual information from human brain activity fMRI has achieved promising performance. However, these decoding models require centralized storage of fMRI data to conduct training, leading to potential privacy security issues. In this paper, we focus on privacy preservation in multi-individual brain visual decoding. To this end, we introduce a novel framework called FedMinds, which utilizes federated learning to protect individuals' privacy during model training. In addition, we deploy individual adapters for each subject, thus allowing personalized visual decoding. We conduct experiments on the authoritative NSD datasets to evaluate the performance of the proposed framework. The results demonstrate that our framework achieves high-precision visual decoding along with privacy protection.

## 1 INTRODUCTION

The human brain is mysterious. While viewing a visual scene, it undergoes significant activity. This brain activity can be captured non-invasively using high spatial-resolution functional magnetic resonance imaging (fMRI) [1, 2]. Based on this, some advanced studies [3, 4, 5, 6, 7, 8, 9] have successfully decoded visual information from these fMRI using data-driven deep learning techniques, achieving promising performance.

Despite the great successes in brain visual decoding [9, 10, 11], existing technologies overlook the privacy issue associated with fMRI data. Specifically, the raw fMRI records not only reflect blood-oxygen-level dependent (BOLD) signals but also contain sensitive information, such as individual identifiers and potential indicators of brain diseases [12]. However, training these deep learning models typically requires centralized data storage [4, 6, 7, 8, 9, 11], which poses a risk of privacy breaches and raises concerns about the practical application of brain visual decoding techniques.

In this paper, we aim to achieve privacy-preserving personalized brain visual decoding. Inspired by a common approach to decentralized machine learning, i.e., federated learning [13], we store each individual's fMRI on individual-specific clients. When jointly training a decoding model, the individuals only need to upload local model weights to the server instead of the raw private fMRI, thus preserving privacy for each participant. Furthermore, considering the huge individual variations [5, 14], we introduce individual-specific local adapters to achieve personalized brain visual decoding. These adapters [15] are only updated during local training but do not participate in model aggregation on the server. We term our solution framework as FedMinds.

To evaluate the performance of our FedMinds, we conducted experiments using the basis brain visual decoding task, i.e., visual category classification. We compare our framework with existing state-of-the-art multi-individual visual classification methods [9] on the authoritative NSD [1] datasets. The experimental results show that our FedMinds can guarantee high-precision brain visual decoding while achieving privacy preservation.

---

* Corresponding author.

Our main contributions can be summarized as:
- Our work is the first to consider privacy preservation in fMRI visual decoding. To this end, we propose a privacy-preserving federated framework for personalized brain visual decoding, called FedMinds.
- We experimentally validated the effectiveness of our framework. It guarantees high-precision visual decoding while achieving privacy preservation.

## 2 RELATED WORKS

### 2.1 Brain Visual Decoding via fMRI

Brain visual decoding via fMRI is an evolving field. Recent studies using deep learning techniques have led to significant progress in this area. Some foundational studies [2, 9] have successfully extracted visual category information from fMRI data, demonstrating the possibility of visual decoding via fMRI. Additionally, with the help of multimodal alignment techniques [16, 17] and pre-trained generative models [18], other pioneering studies [4, 5, 6 ,7, 8, 10] have attempted to reconstruct the visual scenes seen by individuals. Among them, multi-individual visual decoding [5, 9, 10] involves centralized data storage, leading to the potential risk of privacy leakage. However, existing studies have focused on decoding performance while overlooking privacy security issues.

### 2.2 Brain Visual Decoding via fMRI

Federated learning [13, 19, 20] is a common approach to decentralized machine learning. It assumes that multiple clients and a server collaborate to train a model. Since the data used for model training is dispersed across clients and is not visible to other clients or the server, this makes federated learning naturally privacy-preserving. The naive algorithm, i.e. FedAvg [13], involves multiple independent federated communications between clients and the server. In each communication, the server first broadcasts the global model weights to the participating clients. Subsequently, each client loads the weights and trains with local private data and then uploads the updated local model weights back to the server. Finally, the server aggregates the local weights to update the global model.

## 3 METHODS

As shown in Figure 1, our FedMinds achieves privacy-preserving by a FedAvg-based framework. In addition, we employ local models with aggregable networks (yellow) and unaggregatable adapters (green) for each client to achieve individual-specific brain visual decoding. The detailed designs are illustrated as follows.

### 3.1 Federated Learning for Privacy-Preserving Brain Decoding

We assume that there are $N$ individuals participating in the model training for brain visual decoding. The fMRI data of each individual is stored on the individual-specific local client, denoted as $D_n = \{(f_k, l_k)\}_{k=1}^{|D_n|}, n = 1, \ldots, N$, where $f_k$ denotes $k$-th fMRI, $l_k$ denotes the corresponding label of visual category in the form of one-hot codes, and $|D_n|$ denotes the number of fMRI-image pairs provided by the $n$-th individual. The model $M_n(\cdot)$ employed for each individual consists of two parts: aggregable networks $G_n$ and unaggregated adapters $A_n$, respectively parameterized by $\omega_{G_n}$ and $\omega_{A_n}$.

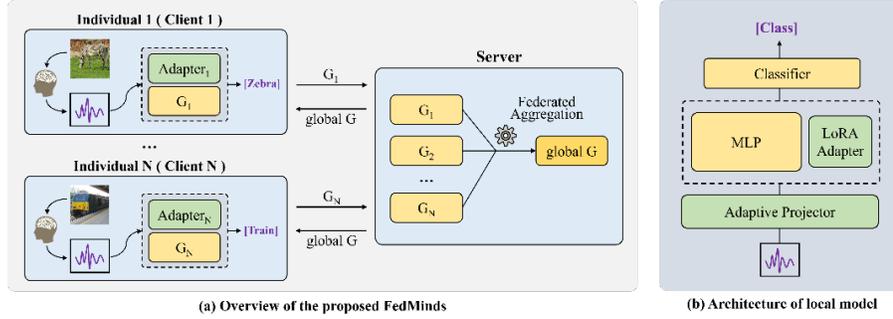

Figure 1: Illustration of our methods for privacy-preserving personalized brain visual decoding.

In each federated communication round, $C$ individuals are selected to participate in local training, where $C \subset \{1, \ldots, N\}$. Each selected individual will update the local model using the supervision of cross-entropy (CE) loss:

$$\mathcal{L}_c = \mathbb{E}_{k \sim D_c}\left[\text{CE}\left(M_c(f_k; \omega_{G_c}, \omega_{A_c}), l_K\right)\right], c \in C. \quad (1)$$

After local training, each participating individual only uploads the aggregable networks $G_c$ to the server. The server then performs model aggregation to obtain the global aggregable network $G_{global}$:

$$\omega_{G_{global}} = \sum_{c \in C} \frac{|D_c|}{|D|} \cdot \omega_{G_c}, D = \bigcup_{c \in C} D_c. \quad (2)$$

Finally, the server broadcasts the global parameters $\omega_{G_{global}}$ to all individuals.

### 3.2 Local Adapters for Personalized Brain Decoding

For each individual, we configure two types of adapters: the adaptive projector and LoRA adapters.

**Adaptive Projector**. Even when seeing the same visual scene, there are differences in brain activity patterns across individuals. Therefore, the adaptive projector serves to embed fMRI into the same (pattern) latent space. The adaptive projector is implemented using ridge regression, commonly used in neuroscience, and needs to be updated throughout local training until the convergence of federated learning.

**LoRA Adapters**. The LoRA adapters are low-rank linear networks attached to an aggregable MLP. Initially, the LoRA adapters are specially initialized so that their outputs are zero, ensuring they do not interfere with the MLP's output. These LoRA adapters remain inactive until the aggregable MLP converges. Once the MLP has converged, both the MLP and the individual's Adaptive Projector are fixed, and the LoRA adapters begin to update. During this phase, the aggregable classifier continues federated training. This above training strategy allows the LoRA adapters to perform customized fine-tuning on the converged MLP, thus enabling personalized decoding.

## 4 EXPERIMENTS

### 4.1 Dataset

The Natural Scenes Dataset is a massive 7T neuroscience dataset. Throughout the NSD experiment, participants were presented with images. Each image has multiple labels from 80 categories. For each subject, the training set comprises 8859 distinct visual stimuli and 24980 fMRI (viewing each image 1-3 times) and the test set contains 982 visual stimuli and 2770 fMRI. The preprocessing aligns fMRI data to the *fsaverage7* template.

Table 1: Experiment results. Baselines are quoted from [25].

| Methods | Privacy | mAP↑ | AUC↑ | Hamming↓ |
|---|---|---|---|---|
| Classifier | × | 0.156 | 0.755 | 0.038 |
| EMB | × | 0.220 | 0.825 | 0.035 |
| CLIP-MUSED | × | 0.258 | 0.877 | 0.030 |
| FedMinds (**ours**) | √ | **0.273** | **0.905** | **0.027** |

Table 2: Results of ablation experiments.

| Methods | mAP↑ | AUC↑ | Hamming↓ |
|---|---|---|---|
| FedMinds | **0.273** | **0.905** | **0.027** |
| w/o Adaptive Projector | 0.134 | 0.801 | 0.038 |
| w/o LoRA Adapter | 0.212 | 0.839 | 0.034 |

### 4.2 Baselines and Evaluation Metrics

We compare with methods for multi-individual brain visual decoding, including simple classifier, EMB [21], and CLIP-MUSED [9]. All of these existing methods store fMRI data centrally and do not protect privacy. Due to NSD classification being a multi-label classification task, we evaluate performance using mean average precision (mAP), the area size under the receiver operating characteristic curve (AUC), and Hamming distance.

### 4.3 Implementation details

For Adaptive Projector (one part of $A_n$) and MLP $G_n$, we adopt the same architecture as MindEye [4]. The entire model has 312M parameters. The LoRA adapters (the other part of $A_n$) are attached to MLP and their ranks are set to 8. We federated train Adaptive Projector, MLP, and classifier for 100 federated communication rounds. Local epoch is 10 and learning rate is 3e-4. Subsequently, we federated train LoRA adapters and classifier for 10 communication rounds and other settings remain the same. Our experiments were conducted using Pytorch and NVIDIA RTX 4090 GPUs.

### 4.4 Results and Analysis

As shown in Table 1, our FedMinds perform better than existing multi-individual brain visual decoding methods on the NSD classification task. Moreover, we successfully protect the privacy of fMRI data. We are surprised to find that even with the data isolation caused by the federated learning framework, the decoding performance of FedMinds exceeds the existing baselines. We attribute this performance increase to our personalized design, i.e., the introduction of Adaptive Projector and LoRA Adapters. These personalized network designs and our training strategy make the decoding models of different individuals heterogeneous, allowing for communal decoding models with individual-specific parameters that can be specifically fine-tuned to an individual's unique brain patterns. The learned brain patterns then feed back into the individual's decoding process. It should be clarified that our FedMinds brings additional computational overhead due to the introduction of privacy protection because the convergence of federated learning is slower than that of centralized learning. Specifically, we require 4.9× the computational overhead of centralized training to achieve similar performance to CLIP-MUSED. However, the extra overhead is worth because we achieve protection for extremely private brain data.

### 4.5 Ablation Studies

We conduct ablation experiments to explore the role of personalized design as discussed in the above analysis. We first ablate the Adaptive Projectors, replacing them with aggregable ridge regressions, and conduct training while ensuring that all else is equal. In another ablation exploration, we ablate the LoRA adapter, completely removing these part networks and their corresponding training. The results of the ablation experiments are shown in Table 2. It can be seen that two types of adapters are useful. In particular, the Adaptive Projector is a key factor in mitigating the performance.

## 5 CONCLUSION

We propose a privacy-preserving personalized brain visual decoding framework called FedMinds. It uses Federated Learning to protect the privacy of fMRI data while achieving personalized decoding using individual-specific adapters. Our experimental results demonstrate that FedMinds performs high-precision visual decoding while protecting privacy.

## ACKNOWLEDGMENTS

This work is supported by the National Key Research and Development Program of China (No. 2022YFB3104700), the National Natural Science Foundation of China (No. 61976158, No. 62376198).

## REFERENCES

[1] Emily J Allen, Ghislain St-Yves, Yihan Wu, Jesse L Breedlove, Jacob S Prince, Logan T Dowdle, Matthias Nau, Brad Caron, Franco Pestilli, et al. 2022. A massive 7T fMRI dataset to bridge cognitive neuroscience and artificial intelligence. Nature neuroscience 25, 1 (2022), 116–126.

[2] Tomoyasu Horikawa and Yukiyasu Kamitani. 2017. Generic decoding of seen and imagined objects using hierarchical visual features. Nature communications 8, 1 (2017), 15037.

[3] Thomas Naselaris, Kendrick N Kay, Shinji Nishimoto, and Jack L Gallant. 2011. Encoding and decoding in fMRI. Neuroimage 56, 2 (2011), 400–410.

[4] Paul Scotti, Atmadeep Banerjee, Jimmie Goode, Stepan Shabalin, Alex Nguyen, Aidan Dempster, Nathalie Verlinde, Elad Yundler, David Weisberg, Kenneth Norman, et al. 2024. Reconstructing the mind's eye: fMRI-to-image with contrastive learning and diffusion priors. arXiv preprint arXiv:2305.18274 (2023).

[5] Guangyin Bao, Zixuan Gong, Qi Zhang, Jialei Zhou, Wei Fan, Kun Yi, Usman Naseem, Liang Hu, and Duoqian Miao. 2024. Wills Aligner: A Robust Multi-Subject Brain Representation Learner. arXiv preprint arXiv:2404.13282 (2024).

[6] Zijiao Chen, Jiaxin Qing, Tiange Xiang, Wan Lin Yue, and Juan Helen Zhou. 2023. Seeing Beyond the Brain: Conditional Diffusion Model with Sparse Masked Modeling for Vision Decoding. In IEEE/CVF Conference on Computer Vision and Pattern Recognition. 22710–22720.

[7] Sikun Lin, Thomas Sprague, and Ambuj K Singh. 2022. Mind reader: Reconstructing complex images from brain activities. Advances in Neural Information Processing Systems 35 (2022), 29624–29636.

[8] Yizhuo Lu, Changde Du, Qiongyi Zhou, Dianpeng Wang, and Huiguang He. 2023. MindDiffuser: Controlled Image Reconstruction from Human Brain Activity with Semantic and Structural Diffusion. In Proceedings of the 31st ACM International Conference on Multimedia. 5899–5908.

[9] Qiongyi Zhou, Changde Du, Shengpei Wang, and Huiguang He. 2024. CLIP-MUSED: CLIP-Guided Multi-Subject Visual Neural Information Semantic Decoding. arXiv preprint arXiv:2402.08994 (2024).

[10] Zixuan Gong, Qi Zhang, Guangyin Bao, Lei Zhu, Ke Liu, Liang Hu, and Duoqian Miao. 2024. MindTuner: Cross-Subject Visual Decoding with Visual Fingerprint and Semantic Correction. arXiv preprint arXiv:2404.12630 (2024).

[11] Zixuan Gong, Qi Zhang, Duoqian Miao, Guangyin Bao, and Liang Hu. 2023. Lite-Mind: Towards Efficient and Versatile Brain Representation Network. CoRR abs/2312.03781 (2023).

[12] Stacey A Tovino. 2005. The confidentiality and privacy implications of functional magnetic resonance imaging. Journal of Law, Medicine & Ethics 33, 4 (2005), 844–850.

[13] Brendan McMahan, Eider Moore, Daniel Ramage, Seth Hampson, and Blaise Agüera y Arcas. 2017. Communication-Efficient Learning of Deep Networks from Decentralized Data. In International Conference on Artificial Intelligence and Statistics, Vol. 54. 1273–1282.

[14] David J McGonigle, Alistair M Howseman, Balwindar S Athwal, Karl J Friston, RSJ Frackowiak, and Andrew P Holmes. 2000. Variability in fMRI: an examination of intersession differences. Neuroimage 11, 6 (2000), 708–734.

[15] Edward J Hu, Yelong Shen, Phillip Wallis, Zeyuan Allen-Zhu, Yuanzhi Li, Shean Wang, Lu Wang, and Weizhu Chen. 2021. Lora: Low-rank adaptation of large language models. arXiv preprint arXiv:2106.09685 (2021).

[16] Alec Radford, Jong Wook Kim, Chris Hallacy, Aditya Ramesh, Gabriel Goh, Sandhini Agarwal, Girish Sastry, Amanda Askell, Pamela Mishkin, Jack Clark, et al. 2021. Learning transferable visual models from natural language supervision. In International conference on machine learning. 8748–8763.

[17] Yu Zhang, Qi Zhang, Zixuan Gong, et al. 2024. MLIP: Efficient Multi-Perspective Language-Image Pretraining with Exhaustive Data Utilization. arXiv preprint arXiv:2406.01460 (2024).

[18] Robin Rombach, Andreas Blattmann, Dominik Lorenz, Patrick Esser, and Björn Ommer. 2022. High-resolution image synthesis with latent diffusion models. In Proceedings of the IEEE/CVF conference on computer vision and pattern recognition. 10684–10695.

[19] Zhuojia Wu, Qi Zhang, Duoqian Miao, Kun Yi, Wei Fan, and Liang Hu. 2024. HyDiscGAN: A Hybrid Distributed cGAN for Audio-Visual Privacy Preservation in Multimodal Sentiment Analysis. arXiv preprint arXiv:2404.11938 (2024).

[20] Guangyin Bao, Qi Zhang, Duoqian Miao, Zixuan Gong, and Liang Hu. 2023. Multimodal Federated Learning with Missing Modality via Prototype Mask and Contrast. arXiv preprint arXiv:2312.13508 (2023).

[21] Omar Chehab, Alexandre Défossez, Loiseau Jean-Christophe, Alexandre Gramfort, and Jean-Remi King. 2022. Deep Recurrent Encoder: an end-to-end network to model magnetoencephalography at scale. Neurons, Behavior, Data Analysis, and Theory, 1(2022), 1-24.